# THREE-CELL TRAVELING WAVE SUPERCONDUCTING TEST STRUCTURE

Pavel Avrakhov#, Alexei Kanareykin (Euclid TechLabs, LLC, Solon, Ohio), Sergey Kazakov, Nikolay Solyak, Genfa Wu, Vyacheslav P. Yakovlev (Fermilab, Batavia)

*Abstract*

Use of a superconducting traveling wave accelerating (STWA) structure [1] with a small phase advance per cell rather than a standing wave structure may provide a significant increase of the accelerating gradient in the ILC [2] linac. For the same surface electric and magnetic fields the STWA achieves an accelerating gradient 1.2 larger than TESLA-like standing wave cavities [3]. The STWA allows also longer acceleration cavities, reducing the number of gaps between them [3]. However, the STWA structure requires a SC feedback waveguide to return the few hundreds of MW of circulating RF power from the structure output to the structure input. A test single-cell cavity with feedback was designed, manufactured and successfully tested demonstrating the possibility of a proper processing to achieve a high accelerating gradient [3, 4]. These results open way to take the next step of the TW SC cavity development: to build and test a traveling-wave three-cell cavity with a feedback waveguide. The latest results of the single-cell cavity tests are discussed as well as the design of the test 3-cell TW cavity.

## INTRODUCTION

The main goal of that work is to study traveling wave superconducting (SC) accelerator concepts to increase the accelerating gradient of a SC structure and reduce the length of the accelerator [1, 3-5]. The STWA allows the increase in the accelerating gradient larger by a factor of 1.2÷1.4 over than that of the TESLA-like designs with a new shape: low-loss [6] or reentrant [7].

The first approach of the traveling wave (TW) SC cavity was, as usual, a single-cell cavity. The single-cell cavity with feedback (see Fig. 1) had to demonstrate the feasibility of the STWA structure [4, 5].

The 1-cell model having the same shape as the regular cell of the full-sized STWA structure allows understanding of the problems of mechanical manufacturing, assembly and welding of this geometry as well as the surface processing issues.

Recently, the first 1-cell model of a TW cavity has been developed and tested [3, 4]. Two test single cell cavities* were manufactured at Advanced Energy System, Inc. (AES) and processed in Argonne and Fermi National Labs. The most critical point was to demonstrate the possibility of a proper processing to achieve a high accelerating gradient. In spite of the reduced processing (without electropolishing) the first results of the high gradient tests were very encouraging. It should be noted the 1-cell model could be tested only in standing wave regime. The test results are presented in Figures 2 and 3.

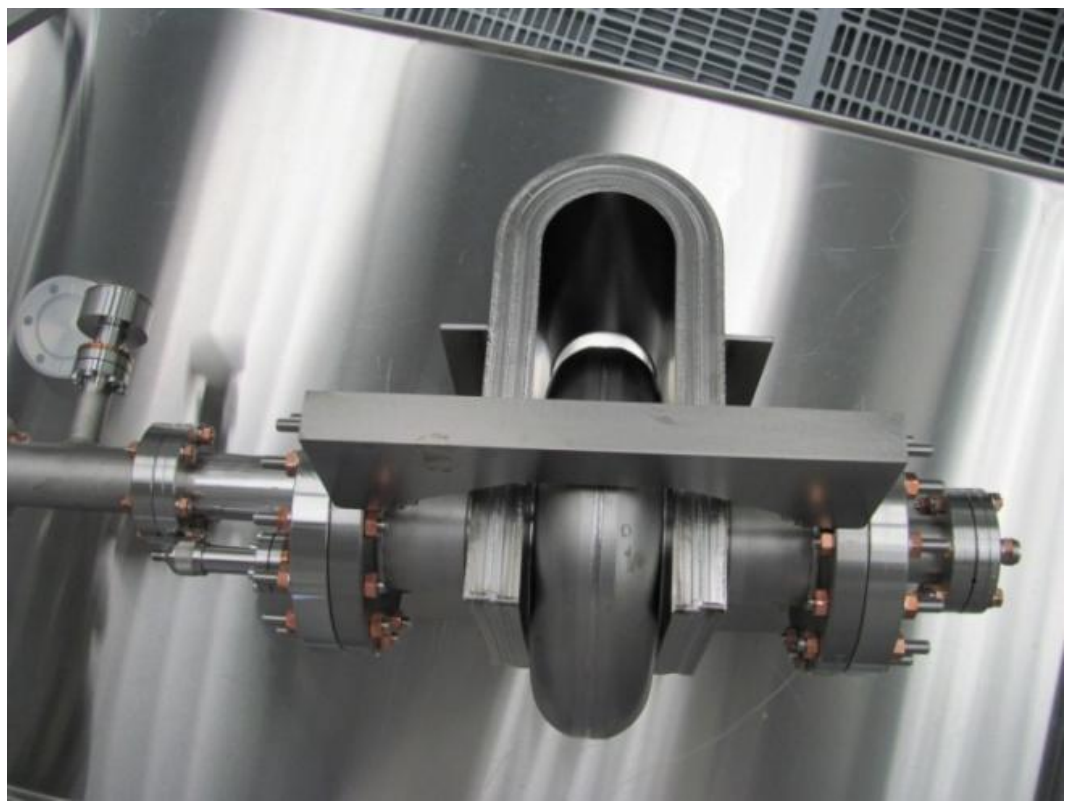

Figure 1. The single-cell model of a traveling wave cavity assembled for evacuation.

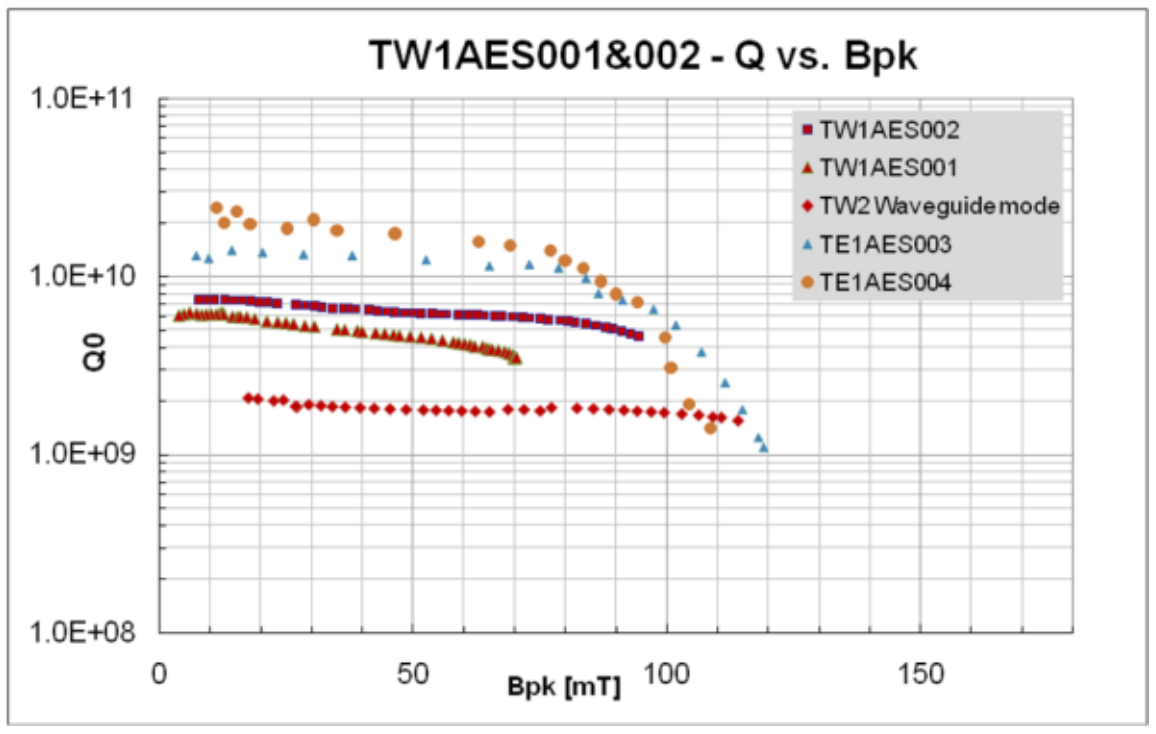

Figure 2. Quality factor vs. surface magnetic field for 1-cell TW cavities (TW1AES001, TW1AES002) in comparison to TESLA-like TE1AES 1-cell cavities.

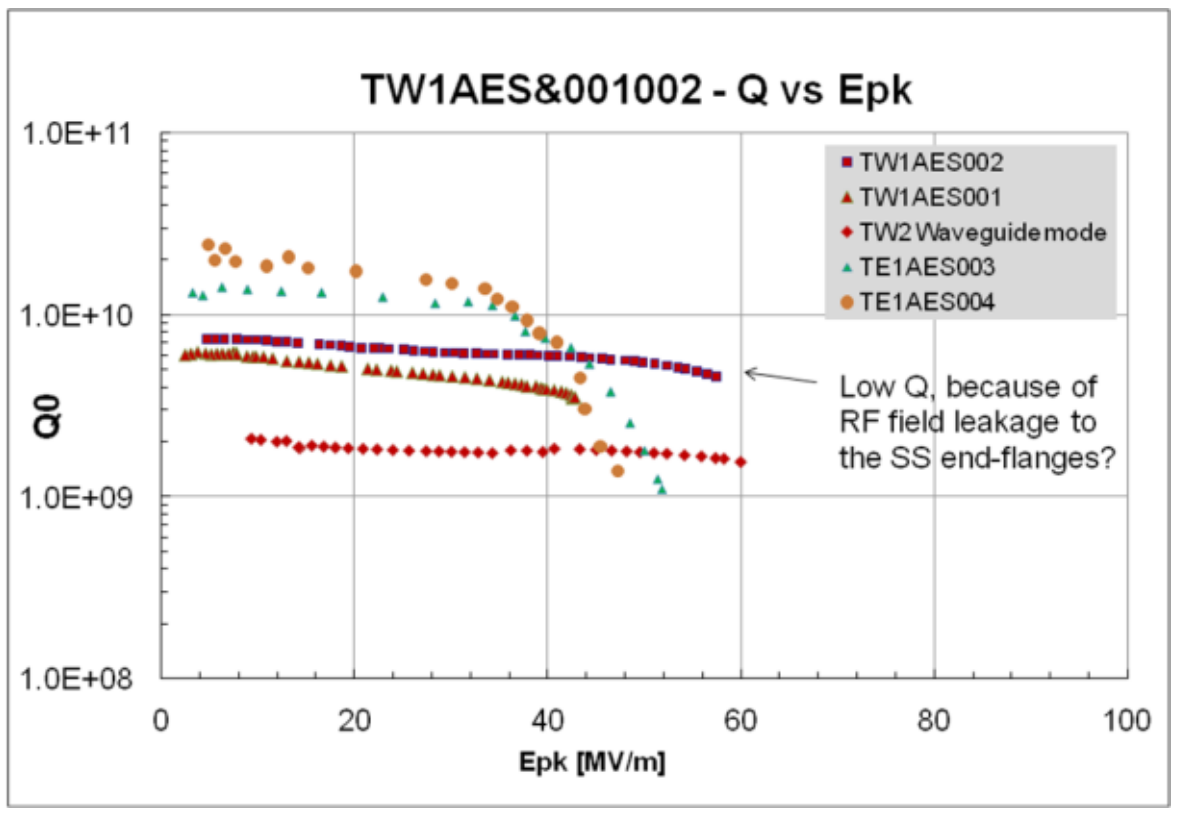

Figure 3. Quality factor vs. surface electric field for 1-cell TW cavities (TW1AES001, TW1AES002) in comparison to TESLA-like TE1AES 1-cell cavities.

*Work supported by US Department of Energy
# p.avrakhov@euclidtechlabs.com

The surface electric field in Figure 3 suggested TW1AES002 reached the equivalent of 31 MV/m of the TESLA-shaped cavity with no field emission. Despite the complex waveguide structure mode with maximum field in the waveguide loop (red diamonds) demonstrated high level magnetic and electric fields.

These results open the way to take the next step of the TW SC cavity development: to build and test a traveling-wave three-cell cavity with a feedback waveguide.

## 3-CELL SUPERCONDUCTING TRAVELING WAVE ACCELERATING STRUCTURE

We consider develop and test a SC TW cavity which contains only one (in the middle of cavity) regular cells. In spite of the fact that 3-cell structure has only one regular 105° cell, there is still the same field distribution and *S*-parameters as for 1 meter 15-cell patterns. S-parameters of the 3-cell TW cavity and field distribution are presented on Fig. 4 and 5 consequently.

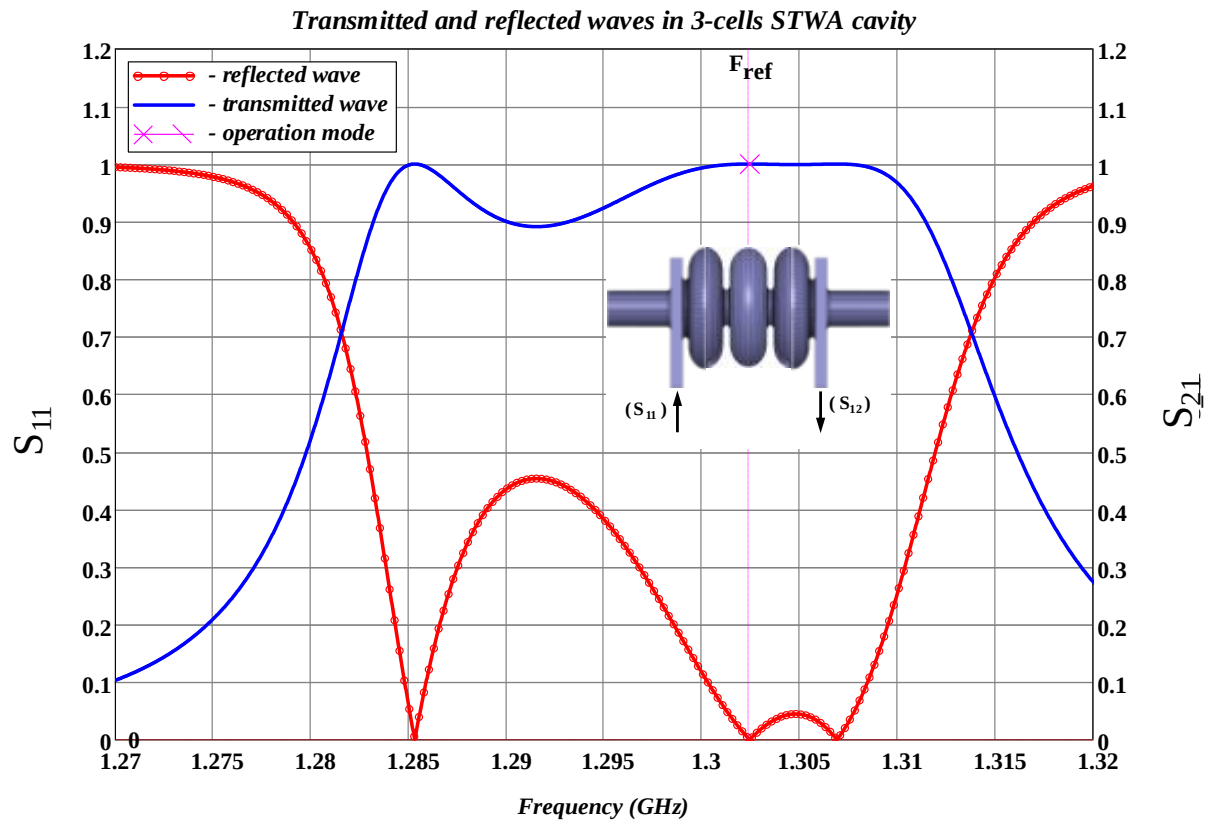

Figure 4. *S*-parameters of the 3-cell TW cavity. and field distribution (down) for operation mode

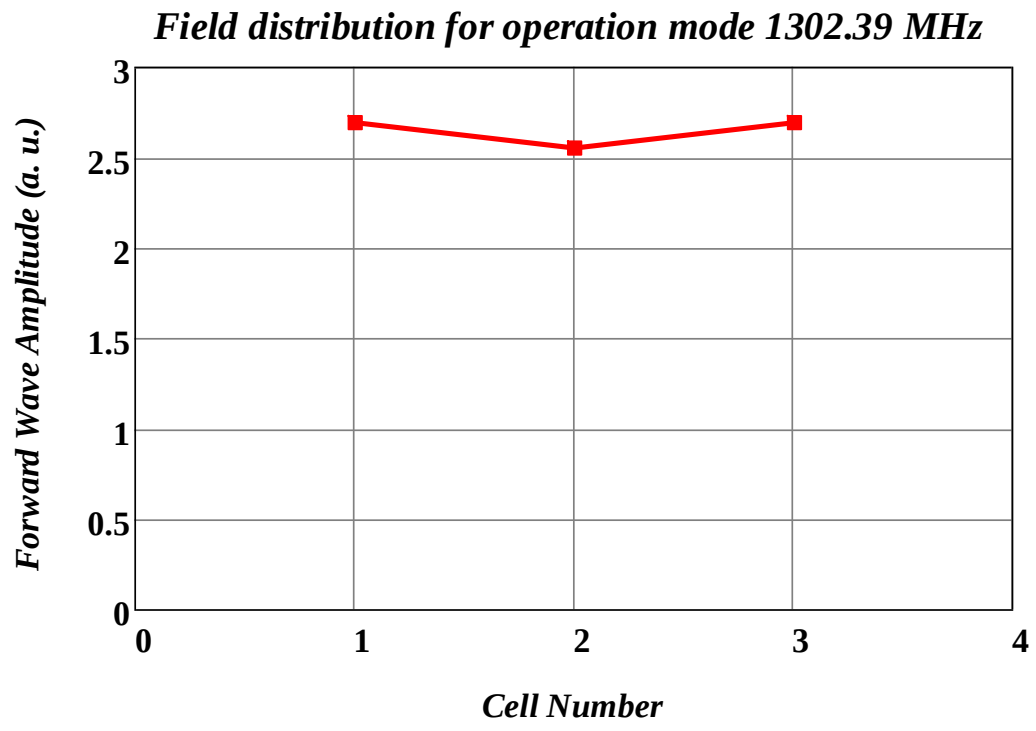

Figure 5. The field distribution for operation mode in the 3-cell TW cavity

The superconducting TW structure involves to utilize the RF power passing through the cavity redirecting it back to the input of the accelerating structure. This scheme of the RF wave circulation in the RF structure requires a feedback loop [5, 8]. The energy out of the RF source goes into the ring resonator (or feedback loop) through the directional coupler to define by the phase relations the correct direction of the RF propagation in the acceleration section. The simplest directional coupler consists of two couplers spaced on a quarter waveguide wavelength. We consider two feeding scheme of the superconducting traveling wave resonator (STWR): 1) two directional couplers in the waveguide loop; 2) two coaxial couplers through the beam pipe (See Fig. 6, 7).

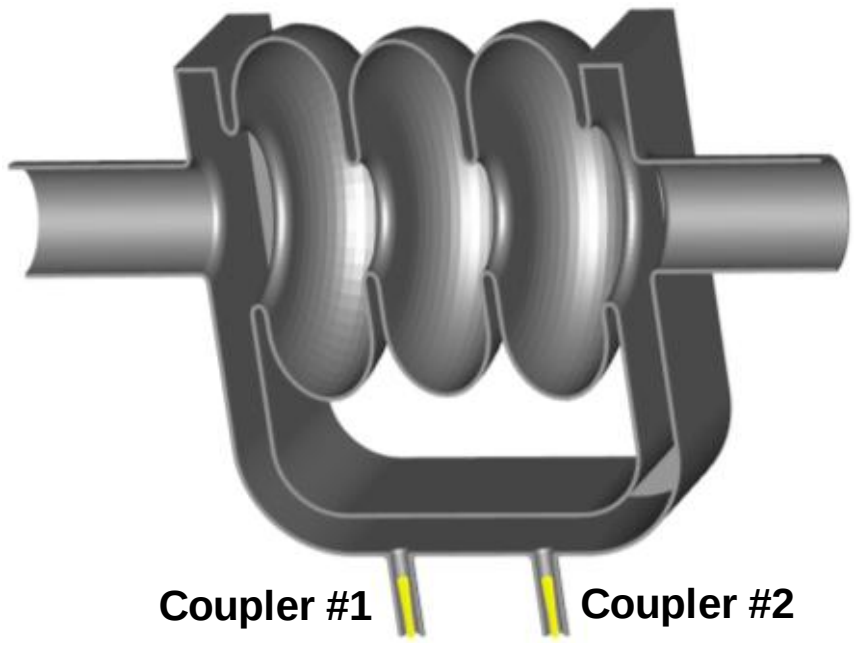

Figure 6. The STWR feeding scheme trough waveguide couplers.

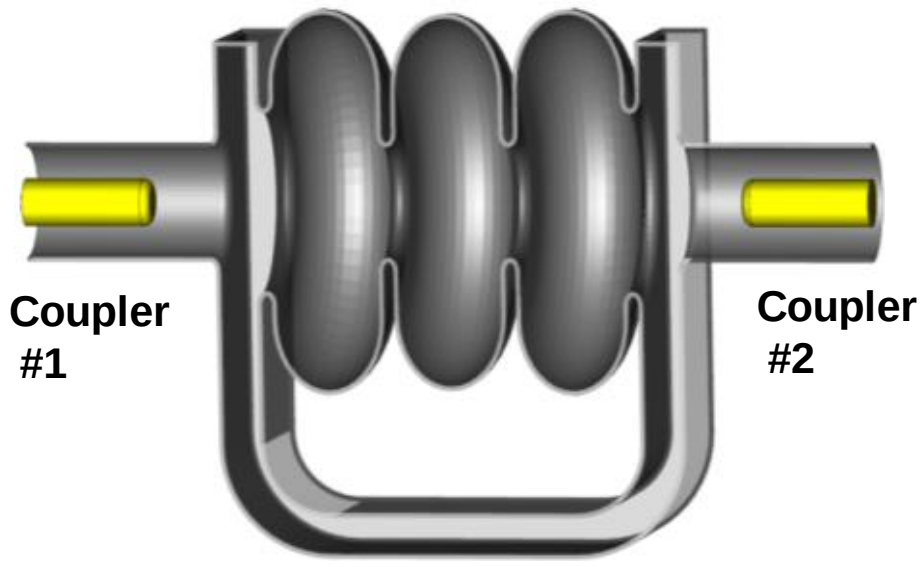

Figure 7. The STWR feeding scheme trough beam pipe.

The first waveguide scheme (Fig. 4) is close to real feeding of the STWR for superconducting linac and this scheme has easier tuning than second one. The second feeding method through beam pipe (Fig. 5) doesn't require additional elements in waveguide loop and allows using standard equipment for single cell test cavities.

Previously developed mathematical tool for STWR design [8] allows to adjust either of the feeding schemes. The Fig. 8 shows tuning results for more complicated case of the beam pipe-fed STWR.

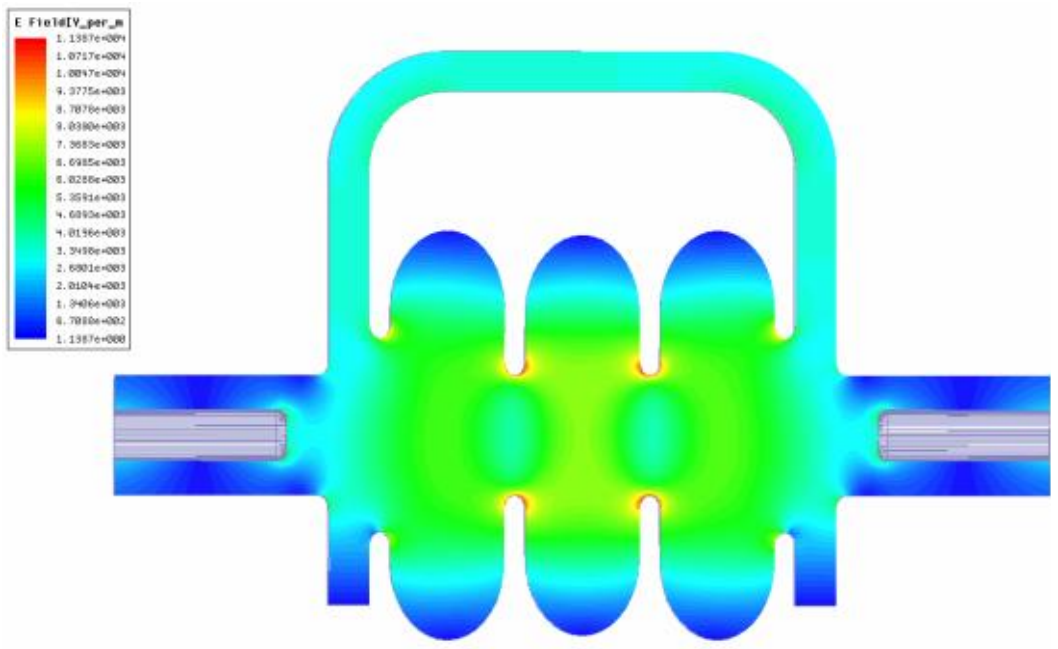

Figure 8. The time-averaged electric field in the fully adjusted STWR.

For the test SC traveling wave resonator the main goal of a tuning procedure is to maintain a high ratio between forward and backward waves into the resonance ring. Sensitivity to the waveguide loop length for STWR with multiplication factor ($P_{forward}/P_{input}$) $K_{mult}$ = 308 illustrates in Fig. 9.

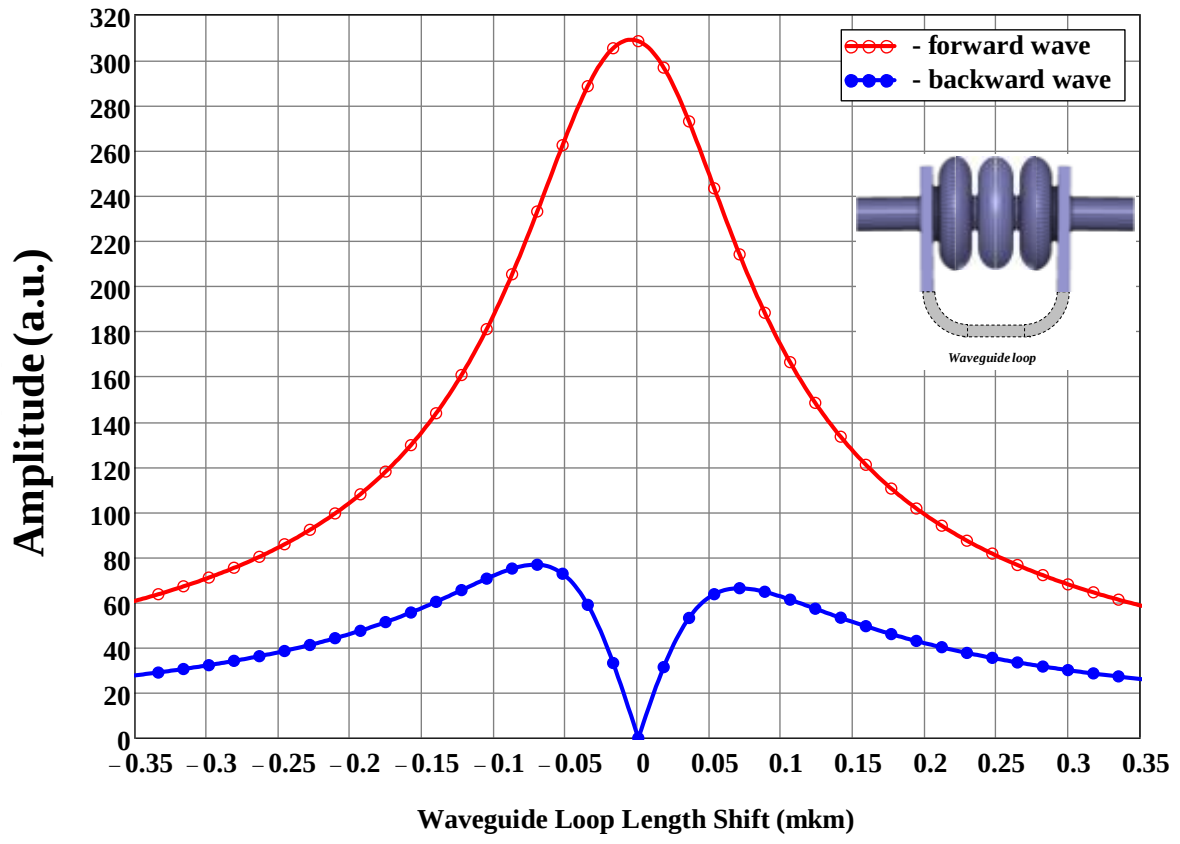

Figure 9. Amplitudes of the forward ($K_{mult}$ = 308) and backward waves vs. waveguide loop length of the 3-cell STWR.

Apparently, the tuning requires extra elements in the resonant ring because of high sensitivity (~$10^{-7}$) of main STWR parameters to the ring and cavity dimensions. Frequency stability and mechanical precision of the superconducting ring with accelerating TW section are very close to SW accelerating cavity with similar accelerating gradient. Fig. 10 shows available STWR schemes with tuning elements.

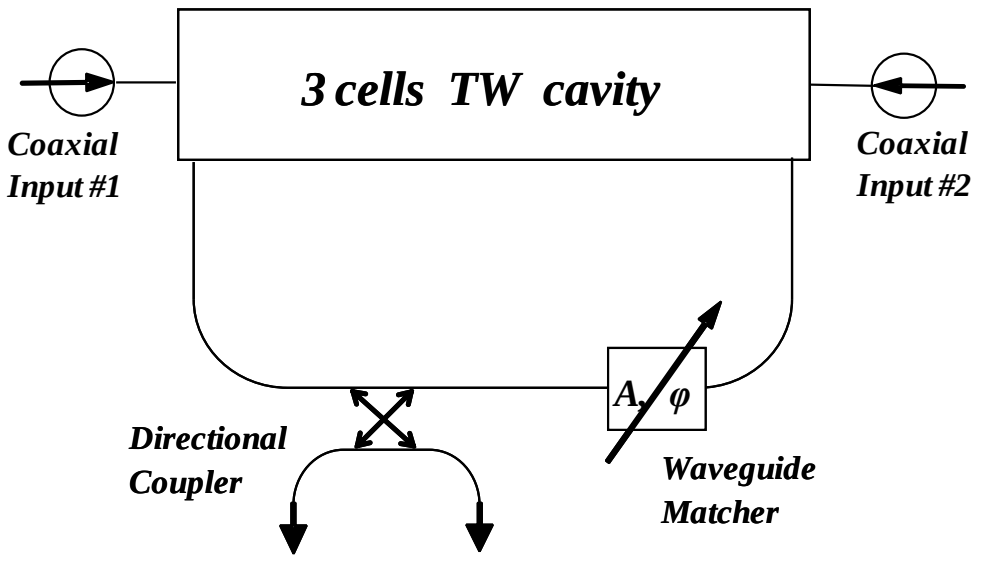

Figure 10. The STWR schemes with extra tuning elements - measuring directional coupler and waveguide matcher.

The important component of traveling wave resonator is a measuring directional coupler. Backward wave into the STWR has to be suppressed at least as 30 dB. It requires the same directivity of the gage. The simple waveguide directional coupler is depictured on Fig. 11. This kind of waveguide directional coupler has pretty low reflection and field enhancement factor due to its placement in the minimum of surface current and RF magnetic field. The central hole of directional coupler is assumed to use for calibration.

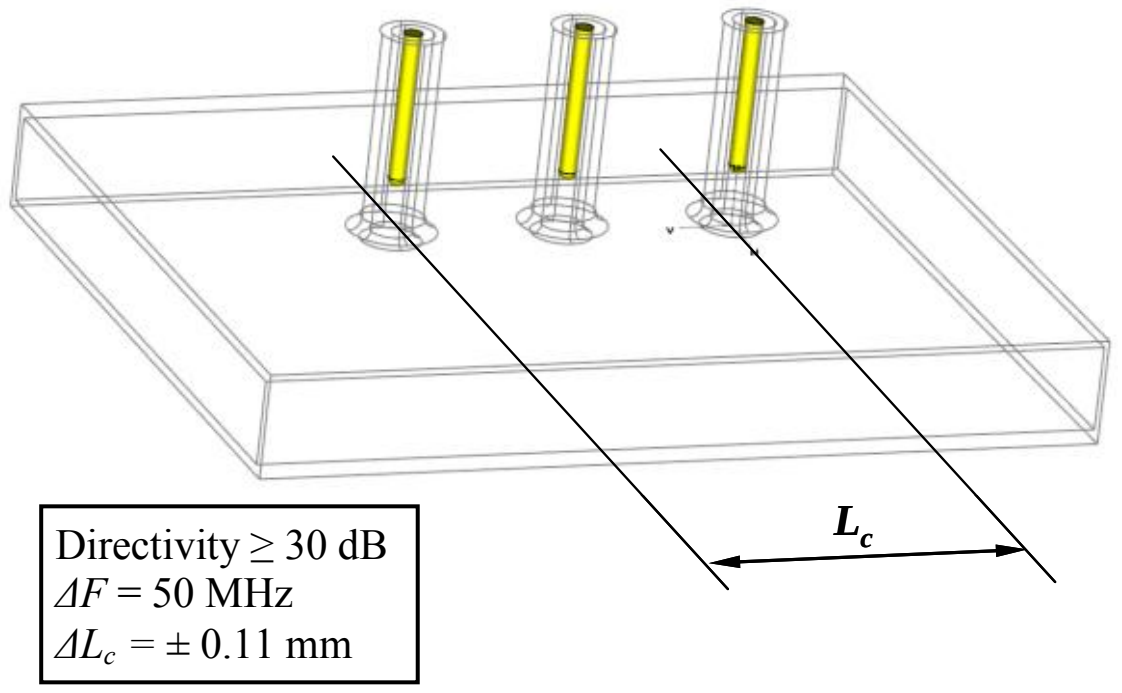

Figure 11. Scheme of the measuring waveguide directional coupler for STWR.

The next resonant ring element is a waveguide matcher. In RF technique it exist many possibilities to change phase length and to arrange control reflection in regular rectangular waveguide.

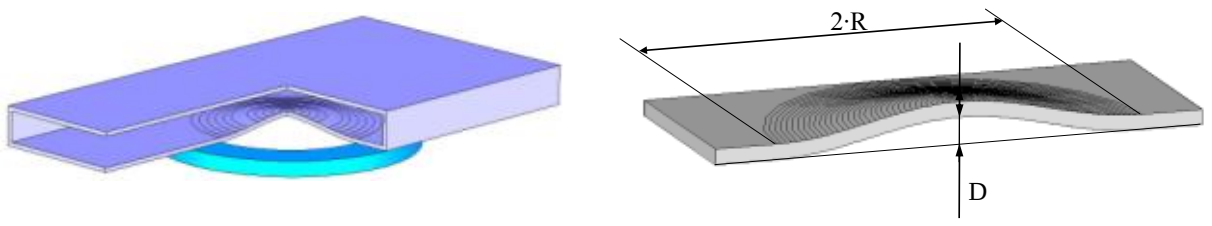

Figure 12. STWR matcher on the wide wall of the waveguide loop

A candidate of waveguide matcher is shown on Fig. 12. A circular fence of the matcher doesn't allow deforming cavity and the waveguide ring shapes.

## CONCLUSION

The successful test results of the single-cell model of STWA structure open the way to take the next step of the traveling wave SC cavity development: to build and test a traveling-wave three-cell cavity with a feedback waveguide. The electromagnetic design of a test three-cell TW cavity is presented. The possible ways of the traveling wave excitation and control are discussed.